\documentclass{aa}

\usepackage[varg]{txfonts}
\usepackage{amssymb}
\usepackage{amsmath}
\usepackage{graphicx}
\usepackage{enumerate}
\usepackage{subeqnarray}
\usepackage{cases}
\usepackage{mathrsfs,amssymb}
\usepackage[usenames]{color}
\usepackage[colorlinks=true,     linkcolor=blue, citecolor=blue, filecolor=blue, urlcolor=blue]{hyperref}

\begin{document}

\title{Secondary cosmic-ray nuclei in the  Galactic halo model with nonlinear Landau damping}

\author{
D.~O.~Chernyshov\inst{\ref{inst1}}\and A.~V.~Ivlev\inst{\ref{inst2}}\and V.~A.~Dogiel\inst{\ref{inst1}} 
}

\institute{
I.~E.~Tamm Theoretical Physics Division of P.~N.~Lebedev Institute of Physics, 119991 Moscow, Russia\\\email{chernyshov@lpi.ru}\label{inst1}
\and
Max-Planck-Institut f\"ur extraterrestrische Physik, 85748 Garching, Germany\\\email{ivlev@mpe.mpg.de}\label{inst2}
}

\date{Received ... / Accepted ...}
       
\abstract{We employed our recent model of the cosmic-ray (CR) halo to compute the Galactic spectra of stable and unstable secondary nuclei. In this model, confinement of the Galactic CRs is entirely determined by the self-generated Alfv\'enic turbulence whose spectrum is controlled by nonlinear Landau damping. We analyzed the physical parameters affecting propagation characteristics of CRs and estimated the best set of free parameters providing accurate description of available observational data. We also show that agreement with observations at lower energies may be further improved by taking into account the effect of ion-neutral damping that operates near the Galactic disk.}

\keywords{(ISM:) cosmic rays -- Galaxy: halo -- turbulence}

\titlerunning{Secondary cosmic-ray nuclei...}
\authorrunning{Chernyshov et al.}

\maketitle
\nolinenumbers

\section{Introduction}

In our recent papers, \citet{dogiel2020} and \citet{chernyshov22}, we developed a self-consistent model of the cosmic-ray halo where confinement of the Galactic cosmic rays (CRs) is entirely determined by the self-generated Alfv\'enic turbulence. Following \citet{ptuskin97}, we assumed that the spectrum of Alfv\'en waves excited via the resonant streaming instability is controlled by nonlinear Landau damping \citep{chernyshov22}. We showed that variations seen in the measured spectral index of CR protons at $\lesssim10^5$~GeV can be accurately explained by the height dependence of the plasma beta -- the key parameter controlling the magnitude of nonlinear Landau damping \citep{Lee1973, miller1991}. 

There have been a number of different physical models proposed to explain the observed variations in the spectral index of protons and heavier nuclei in this energy range. In particular, these include scenarios of the spectral break in the CR injection spectra as well as of local sources at low and high energies \citep{Vladimirov2012, Fornieri2021}; different mechanisms for CR confinement at different energies, which are associated with scattering on both self-generated and pre-existing MHD turbulence \citep{Aloisio2015, evoli18, Fornieri2021a, Kempski2022}; a local supernova remnant as an additional CR source \citep[e.g.,][]{erlykin15, Yang2019, Liu2019}; and CR reacceleration at a stellar bow shock \citep{Malkov2021, Malkov2022}. Therefore, the aim of the present paper is to explore the ability of our model to reproduce the measured spectra of heavier CR nuclei.

Apart from available observational data on primary CRs, additional information about CR transport in the Galactic halo can be derived from the data on secondary nuclei, which include isotopes of Li, Be, and B. The abundance of these nuclei in CRs is much greater than in the interstellar medium, and they are believed to be produced by nuclear interactions of primary CRs with background gas \citep[see, e.g., the review by][]{stron07}. The boron-to-carbon ratio is normally employed to test CR propagation models, since boron is mostly produced from carbon and cross-sections for boron production are well known.

Propagation of CRs can be further constrained by studying unstable CR isotopes. \citet{hayakawa58} suggested that the abundance of unstable $^{10}$Be contains information on how long CRs are confined in the Galactic halo. Since the decay time of the unstable isotope is estimated to be comparable to the confinement time, the abundance ratio $^{10}$Be/$^{9}$Be must be a strong function of the confinement time and thus can be used to derive characteristics of the CR halo. The downside of this approach is that observations require careful separation of the two isotopes, which is a complicated task.

In this paper, we applied our self-consistent model by \citet{chernyshov22} to compute the Galactic spectra of stable and unstable secondary nuclei. We explored the physical parameters affecting propagation characteristics of CRs and estimated the best set of free parameters, which allowed us to accurately describe available observational data.

\section{Governing equations for CR nuclei}

GALPROP\footnote{https://galprop.stanford.edu/} \citep{mosk98} is one of the most sophisticated numerical codes employed to calculate spectra of CR species in the Galaxy. For our purposes, however, GALPROP cannot be directly used due to the following reasons. First, the magnitude and the energy dependence of the diffusion coefficient $D$ in GALPROP are assumed, and $D$ is set to be constant in space -- whereas in \citet{chernyshov22} $D(p,z)$ is a solution of a self-consistent model, resulting in a strong dependence on the vertical coordinate $z$. Second, GALPROP utilizes cylindrical or 3D geometry, while our model is one-dimensional. Therefore, below we numerically solve a set of transport equations for the primary and secondary CR nuclei, using the self-consistent solution for protons from \citet{chernyshov22} and utilizing the nuclear reaction network from GALPROP.

To describe propagation of CR nuclei $i$, we use the one-dimensional transport equation for their spectrum $N_i(p,z)$:
\begin{equation}
\begin{array}{l}
{\displaystyle\frac{\partial N_i}{\partial t} + \frac{N_i}{\tau_{{\rm fr},i}} + \frac{N_i}{\tau_{{\rm dec},i}} +
\frac{\partial}{\partial z} \left(u_{\rm adv}N_i - D_i\frac{\partial N_i}{\partial z}\right)}
\\[.4cm]
{\displaystyle-\frac{\partial}{\partial R}\left(\frac13\frac{du_{\rm adv}}{dz}RN_i -\dot{R}_iN_i\right) = Q_i + q_i}\,,
\end{array}
\label{eq:main_iso}
\end{equation}
where $R = pc/eZ_i$ is the rigidity, related to the momentum $p$ via the atomic number $Z_i$. The term $\dot{R}_i(R) \equiv \dot p_i(R)c/eZ_i< 0$ describes continuous momentum losses due to ionization and Coulomb collisions. The timescales $\tau_{{\rm fr},i}(R)$ and $\tau_{{\rm dec},i}$ are, respectively, the fragmentation time of nuclei $i$ due to their collisions with background gas and the decay time (in case of unstable nuclei). The former is given by $1/\tau_{{\rm fr},i}=n_{\rm H}\sigma_{{\rm fr},i}v_i$, where $n_{\rm H}$ is the gas number density, $v_i$ is the velocity of the nuclei, and $\sigma_{{\rm fr},i}(R)$ is the fragmentation cross section. 

The diffusion coefficient $D_i$ of nuclei $i$ is proportional to the product of the velocity $v_i$ and the mean free path that is a sole function of rigidity (for a given $z$). Therefore, $D_i$ is related to the diffusion coefficient of protons $D_p$ via
\begin{equation}
D_i =\frac{v_i}{v_p} D_p \,,
\label{eq:diff_nuclei}
\end{equation}
while the ratio of $v_i$ to the proton velocity $v_p$ is
\begin{equation}
\frac{v_i}{v_p} = \sqrt{\frac{R^2 + (m_pc^2/e)^2}{R^2 + (m_pc^2A_i/eZ_i)^2}} \,,
\end{equation}
where $A_i$ is the atomic mass of the nuclei. The dependence $D_p(R,z)$, derived from our self-consistent model \citep{chernyshov22}, sets the spatial profile of the proton spectrum in the halo, $N_p(R,z)$ (see Section~\ref{analytical}). The advection velocity $u_{\rm adv}$ is assumed to be equal to the local Alfv\'en velocity $u_{\rm A}$ in the halo, vanishing discontinuously at the disk boundary $z=z_{\rm d}$,
\begin{equation}
u_{\rm adv}(z) = u_{\rm A}(z)\theta(z - z_{\rm d}) \,,
\label{eq:disk_to_halo_conv}
\end{equation}
where $\theta(z)$ is the Heaviside function.

The source terms $Q_i(R,z)$ and $q_i(R)$ describe the production of, respectively, primary CR nuclei and secondary nuclei. The source term $Q_i$ can be approximated by
\begin{equation}
Q_i(R,z) = C_i R^{-\gamma}\delta(z) \,,
\label{eq:prim_CR_source}
\end{equation}
where the spectral index $\gamma$ is the same for all primary nuclei, and $\delta(z)$ is the delta function. Constants $C_i$ are adjusted to fit observational data on primary CR spectra, and $C_i = 0$ for secondary nuclei. The source term $q_i$ has two contributions. One is associated with the fragmentation of primary CRs due to their collisions with background gas: 
\begin{equation}
q_{{\rm fr},i}(R) = n_{\rm H} v_i\sum\limits_{j} \frac{Z_iA_j}{Z_jA_i} \sigma_{ji}(\tilde R) N_j(\tilde R) \,,
\label{eq:sec_term}
\end{equation}
where $\sigma_{ji}$ is the total inclusive cross-section of production of secondary nuclei $i$ from nuclei $j$. The product $\sigma_{ji}N_j$ is evaluated at $\tilde R= \frac{Z_iA_j}{Z_jA_i}R$, which takes into account the fact that nuclear reactions conserve the kinetic energy (momentum) per nucleon, not the rigidity. Similarly, the second contribution due to decay of unstable nuclei is given by
\begin{equation}
q_{{\rm dec},i}(R) = \sum\limits_{j} \frac{Z_iA_j}{Z_jA_i} \,\frac{N_j(\tilde R)}{\tau_{{\rm dec},j}} \,.
\label{eq:sec_term2}
\end{equation}

We use the advection velocity $u_{\rm adv}(z)$ and the proton diffusion coefficient $D_p(R,z)$ from the self-consistent model by \citet{chernyshov22}. The diffusion coefficients $D_i$ for heavier nuclei are calculated from Equation~(\ref{eq:diff_nuclei}). To estimate the momentum loss terms $\dot{R}_i(R)$, the fragmentation cross-sections $\sigma_{{\rm fr},i}(R),$ and the total inclusive cross sections $\sigma_{ji}(R)$, we use the GALPROP code (\texttt{energy\_losses.cc}, \texttt{nucleon\_cs.cc}, and \texttt{decayed\_cross\_sections.cc}, respectively). 

The set of equations (\ref{eq:main_iso}) with the secondary source terms (\ref{eq:sec_term}) and (\ref{eq:sec_term2}) is solved numerically until a stationary solution is reached. To reduce computation time, we take into account only the following primary nuclei: $^4$He, $^{12}$C, $^{14}$N, $^{16}$O, $^{20}$Ne, $^{24}$Mg, $^{28}$Si, $^{32}$S, and $^{56}$Fe; for secondary nuclei, we consider $^{6}$Li, $^{7}$Li, $^{7}$Be, $^{9}$Be, $^{10}$Be, $^{10}$B, $^{11}$B, $^{13}$C, $^{14}$C, $^{15}$N, $^{17}$O, and $^{18}$O.

We note that the unstable nuclei $^7$Be decay via the electron capture, with the lifetime of about 100 days for hydrogen-like ions formed after recombination with the surrounding electrons. Therefore, for our purposes we assume that $^7$Be nuclei decay immediately after recombination. This assumption is well justified, because for these nuclei we are only interested in the kinetic energies below $\sim10^3$~GeV/nucleon. The electron recombination cross-section from GALPROP is added to the fragmentation cross-section of $^7$Be and also used as the production cross-section for the reaction $^7{\rm Be} + e \to\,^7{\rm Li}$.

\section{Analytical estimates for primary and secondary CR spectra}
\label{analytical}
In this section, we briefly summarize predictions of \citet{chernyshov22} for spectra of primary CR nuclei, and we apply these to derive and analyze simple scaling relations for stable and unstable secondary nuclei. Similarly to approaches applied earlier to constrain diffusion parameters in the halo \cite[see, e.g.,][]{maurin01}, this allows us to compare the theoretical predictions with observations and put constrains on our self-consistent model (see Section~\ref{results}).

The one-dimensional model by \citet{chernyshov22} considers two regions: the Galactic disk, located at $0 \leq z \leq z_{\rm d}$ and the Galactic halo at $z > z_{\rm d}$. The advection velocity in the halo equals to the Alfv\'en velocity $u_{\rm A}(z)$, and continuous momentum losses vanish in this region. All sources of primary CRs are located in the disk, where we assume CR diffusion in the vertical direction with the diffusion coefficient $D_{\rm d}$. Also, we suppose that turbulence in the disk is characterized by zero cross-helicity (that is, spectra of waves propagating in both directions are balanced), and, hence, advection is absent.
For simplicity, we neglect the vertical gradient of CR density in the disk. The latter implies the following necessary condition for the fragmentation and decay timescales:
\begin{equation}
\frac{1}{\sqrt{D_{\rm d}\tau_{{\rm fr},i}}} + \frac{1}{\sqrt{D_{\rm d}\tau_{{\rm dec},i}}} \ll \frac{1}{z_{\rm d}} \,.
\label{eq:nograd_condition}
\end{equation}
We point out that this condition is satisfied for values of $D_{\rm d}\sim10^{28}$~cm$^2$/s (typically assumed for GeV particles), and therefore the results presented in Section~\ref{results} do not depend on the particular choice of $D_{\rm d}$.

In \citet{chernyshov22}, we calculated the outflow velocity for protons $u_p(R,z) $ self-consistently; this relates the proton spectrum in the halo $N_p(R,z)$ to the outgoing flux of protons at the disk boundary, $S_p(R)$,
\begin{equation}
N_p(R,z) = \frac{S_p(R)}{u_p(R,z)} \,,
\label{eq:N_simple_stable}
\end{equation}
where
\begin{equation}
u_p = \left(\int\limits_\eta^{\infty} \frac{e^{\eta-\eta_1}d\eta_1}{u_{\rm A}(\eta_1)} \right)^{-1}
\label{eq:u_definition}
\end{equation}
and $\eta(R,z)$ is a dimensionless variable:
\begin{equation}
\eta = \int\limits_{z_{\rm d}}^z \frac{u_{\rm A}}{D_p} dz_1 \,.
\label{eq:eta}
\end{equation}
According to Equations~(3) and (23) in \citet{chernyshov22}, we have $D_p(R,z)\propto -g(z)/[R^2S_p(R)]'$, where $g(z)$ describes the height dependence of nonlinear Landau damping (see Equation~(7) therein), and the prime denotes the derivative with respect to $R$. The function $g(z)$ plays a key role in shaping the self-consistent proton spectrum and determining the energy dependence of the halo size \citep[see][for details]{chernyshov22}: $g$ and hence $D_p$ decrease with $z$ while $u_{\rm A}$ increases. Thus, the upper halo boundary is set where the advection contribution $u_{\rm A}N_p$ to the outgoing proton flux becomes equal to the diffusion contribution $-D_p\partial N_p/\partial z$, which is equivalent to the condition $\eta(R,z)\sim1$.

The outflow velocity $u_i(R,z)$ for heavier primary species only deviates from $u_p(R,z)$  due to the factor $v_i/v_p$ in the diffusion coefficient, Equation~(\ref{eq:diff_nuclei}). For the estimates below, we assume relativistic particles with $v_i \approx c \approx v_p$; it is thus safe to set $D_i\approx D_p$ and $u_i \approx u_p$.

Furthermore, continuous momentum losses operating in the disk are negligible for protons with the kinetic energy above $\sim0.5$~GeV \citep{chernyshov22}, and the outgoing flux in this case is equal to the accumulated source of protons in the disk (see Equation~(20) therein; $S_p(R)\approx\frac12C_pR^{-\gamma}$ in our notations). The threshold rigidity $R_{\rm th}$, above which the losses are negligible for heavier nuclei, scales as $R_{\rm th}\propto (A_i^2/Z_i)^{1/3}$ for sub-relativistic particles and $R_{\rm th}\propto Z_i$ in the ultra-relativistic limit. In our analysis (see Section~\ref{results}), we are interested in rigidity values of up to $\sim10^5$~GV, and thus the effect of continuous losses can be reasonably neglected.

We first consider stable secondary nuclei. By integrating Equation~(\ref{eq:main_iso}) across the disk and matching the resulting flux at the disk boundary with the outgoing flux according to Equation~(\ref{eq:N_simple_stable}), we can estimate the secondary spectra $N_0(R)=N(R,z_{\rm d})$ in the disk:
\begin{equation}
N_0(R)\approx\frac{q_{\rm fr}(R)z_{\rm d}}{u_0(R)+u_{\rm fr}(R)}\,
\label{eq:sec_to_prim_ratio}
\end{equation}
(here and below, nuclei index $i$ is omitted for brevity). The numerator, which is the integral over the secondary source term, is proportional to the vertical column density of hydrogen atoms in the disk, $\mathcal{N}_{\rm H}= n_{\rm H}z_{\rm d}$; the denominator is a sum of the outflow velocity at the disk boundary, $u_0(R) = u_p(R,z_{\rm d})$, and the velocity scale $u_{\rm fr}(R) = \mathcal{N}_{\rm H}\sigma_{\rm fr}(R)v\equiv z_{\rm d}/\tau_{\rm fr}(R)$ is associated with fragmentation of the secondary nuclei. After substituting Equation~(\ref{eq:sec_term}) for $q_{\rm fr}(R)$, the obtained expression allows us to estimate abundance ratios of secondary-to-primary nuclei and compare them with available measurements. Keeping in mind that $q_{\rm fr}(R)$ and $u_{\rm fr}(R)$ have similar dependencies on rigidity (since both scale with $\sigma_{ji}(R)$, similar for different nuclei), whereas observations generally exhibit a strong dependence of the secondary to primary ratios on $R$, we conclude that $u_0$ must be substantially larger than $u_{\rm fr}$. Hence, for assumed values of $\mathcal{N}_{\rm H}$, observations yield the dependence $u_0(R)$ that can be compared with the prediction of \citet{chernyshov22}. 

Equation (\ref{eq:sec_to_prim_ratio}) cannot be used for unstable nuclei, since their flux in the halo is not conserved, and thus Equation~(\ref{eq:N_simple_stable}) is no longer applicable. To take into account the decay, one can introduce the ``survival probability'' $P(R,z)$, which relates the spectrum of unstable nuclei $N^*(R,z)$ in the halo to their outgoing flux $S^*(R)$ at the disk boundary:
\begin{equation}
N^*(R,z) = P(R,z)\frac{S^*(R)}{u(R,z)}\,.
\label{eq:surv_def}
\end{equation}
The outflow velocity $u(R,z)$ is given by Equation~(\ref{eq:u_definition}), while the unknown value of $S^*(R)$ is calculated from the matching condition for the flux at $z = z_{\rm d}$.

Substituting Equation~(\ref{eq:surv_def}) in Equation~(\ref{eq:main_iso}) leads to the following equation for $P(R,z)$ in the halo:
\begin{equation}
\frac{d}{dz}\left(\frac{D(R,z)}{u(R,z)}\frac{dP}{dz}\right) = \frac{dP}{dz}+\frac{P}{u(R,z)\tau_{\rm dec}} \,.
\label{eq:surv_eq_1}
\end{equation}
By introducing a new dimensionless variable $x = \int_{z_{\rm d}}^z (u/D)dz_1$, we obtain
\begin{equation}
\frac{d^2P}{dx^2} = \frac{dP}{dx} + \frac{DP}{u^2\tau_{\rm dec}} \,.
\end{equation}
Then, by assuming $P = \tilde{P}e^x$ and $y = \int_0^x e^{-x_1}dx_1$, we reduce it in the Shr\"odinger equation,
\begin{equation}
\frac{d^2\tilde{P}}{dy^2} = \frac{De^{2x}}{u^2\tau_{\rm dec}}\tilde{P} \,,
\label{eq:shrodinger}
\end{equation}
which can be solved using the WKB method. The necessary condition for the WKB method to be applied is that the ``potential'' on the rhs of Equation~(\ref{eq:shrodinger}) varies sufficiently slowly with $y$, that is,
\begin{equation}
\left|\frac{d}{dy}\left(\frac{De^{2x}}{u^2\tau_{\rm dec}}\right)\right| \ll \frac{D^{3/2}e^{3x}}{(u^2\tau_{\rm dec})^{3/2}} \,,
\end{equation}
which leads to
\begin{equation}
\left|\frac{1}{2D}\frac{\partial D}{\partial z} + \frac{u_{\rm A}}{D}\right|^{-1} \gg \sqrt{D\tau_{\rm dec}} \,.
\label{eq:WKB_sufficient_condition}
\end{equation}
The lhs of this condition can be considered as characteristic halo size; according to \citet{dogiel2020}, the upper halo boundary either forms due to a sudden increase in the diffusion coefficient -which reflects a transition to CR free-streaming in Equation~(\ref{eq:main_iso})- or it is set where the advection contribution $u_{\rm A}N$ to the outgoing flux starts dominating. The rhs is roughly the characteristic height that unstable nuclei can reach. Thus, the necessary condition for the WKB approximation to be applicable requires $\tau_{\rm dec}$ to be small enough that unstable nuclei are not reaching the halo boundary.

Substituting the leading term of the WKB solution for $\tilde P(y)$ and keeping in mind that $P(R,z)$ should be finite at $z = \infty$ (as sources of unstable nuclei are absent in the halo), we obtain
\begin{equation}
P \approx A\left(\frac{u^2\tau_{\rm dec}}{D}\right)^{1/4}\exp\left[\;\int\limits_{z_{\rm d}}^z \left(\frac{u}{D} - \frac{1}{\sqrt{D\tau_{\rm dec}}}\right) dz_1\right] \,,
\end{equation}
where the constant $A$ is calculated from the boundary condition at $z = z_{\rm d}$, 
\begin{equation}
\left.\left(u_{\rm A}N^* - D\frac{\partial N^*}{\partial z}\right)\right|_{z = z_{\rm d}} = S^* \,.
\end{equation}
This yields the following expression for $P_0(R) \equiv P(R,z_{\rm d})$:
\begin{equation}
P_0 \approx 
\sqrt{\frac{u_0^2\tau_{\rm dec}}{D_0}}\,,
\label{eq:prob_expr}
\end{equation}
where $D_0(R) \equiv D(R,z_{\rm d})$, $u_0(R) \equiv u(R,z_{\rm d})$, and small terms are omitted according to Equation~(\ref{eq:WKB_sufficient_condition}). We note that $P_0$ is always much smaller than unity (and thus is a function of $D_0$ and $u_0$) if the WKB condition~(\ref{eq:WKB_sufficient_condition}) is satisfied in the advection-dominated regime.

Spectra of unstable nuclei in the disk, $N_0^*(R)$, are estimated using a matching condition for their flux, similar to that used to derive Equation~(\ref{eq:sec_to_prim_ratio}). Assuming that condition~(\ref{eq:nograd_condition}) is satisfied, we integrate Equation~(\ref{eq:main_iso}) across the disk and, employing Equation~(\ref{eq:surv_def}), obtain
\begin{equation}
N_0^*(R) = \frac{P_0(R)q(R)z_{\rm d}}{u_0(R) + P_0(R)\left[u_{\rm fr}(R) +z_{\rm d}/\tau_{\rm dec}\right]} \,,
\label{eq:S0_unstable}
\end{equation}
where $q=q_{\rm fr}+q_{\rm dec}$ according to Equations~(\ref{eq:sec_term}) and (\ref{eq:sec_term2}). As discussed after Equation~(\ref{eq:sec_to_prim_ratio}), observational data suggest that $u_0\gtrsim u_{\rm fr}$, and therefore the term $u_{\rm fr}\equiv z_{\rm d}/\tau_{\rm fr}$ can also be dropped in Equation~(\ref{eq:S0_unstable}). Then two limits can be considered. One is when $\tau_{\rm dec}$ is too small and the contribution of the second term in the brackets, $P_0z_{\rm d}/ \tau_{\rm dec}$, is much larger than $u_0$. In this ``useless'' case, $N_0^*(R)$ depends neither on $u_0$ nor on $P_0$; therefore, it does not contain information on the diffusion coefficient. In the opposite limit, if
\begin{equation}
P_0\frac{z_{\rm d}}{\tau_{\rm dec}} \ll u_0 \,
\label{eq:thin_unstable_disk}
\end{equation}
and condition~(\ref{eq:WKB_sufficient_condition}) is satisfied, abundance ratios of unstable to stable secondary nuclei derived from Equations~(\ref{eq:sec_to_prim_ratio}) and (\ref{eq:S0_unstable}) are proportional to $P_0(R)\propto u_0(R)/\sqrt{D_0(R)}$. Hence, the results can be compared with a rigidity dependence deduced from observational data, which allows us to verify the model prediction of \citet{chernyshov22}. Comparison with beryllium and boron measurements are particularly convenient here, as boron is partially produced due to decay of $^{10}$Be. We point out that if $\tau_{\rm dec}\ll\tau_{\rm fr}$, then condition~(\ref{eq:thin_unstable_disk}) is equivalent to the initial necessary condition~(\ref{eq:nograd_condition}) of homogeneity within the disk.

The main results of the simple analysis presented in this section can be summarized as follows.
\begin{itemize}
\item Measured abundance ratios of stable secondary-to-primary nuclei can be used to verify the expression for the outflow velocity $u_0(R)$, which is derived from the self-consistent model by \citet{chernyshov22}. Below, we employ the B/C ratio for this purpose.
\item Measured abundance ratios of unstable-to-stable secondary nuclei can be used to verify the derived diffusion coefficient $D_0(R)$ and, thus, the predicted size of the halo. Such analysis is possible if the decay timescale $\tau_{\rm dec}$ of unstable nuclei is within a certain range, satisfying conditions~(\ref{eq:WKB_sufficient_condition}) and (\ref{eq:thin_unstable_disk}). Here, one can either use the Be/B ratio, since Be includes unstable $^{10}$Be, or the $^{10}$Be/$^9$Be ratio directly if observational data are available.
\end{itemize}

\section{Discussion}
\label{results}

In this section, we present and discuss results of numerical solution of Equations (\ref{eq:main_iso}), and compare this with available observational data.

\subsection{Numerical results and comparison with observations}\label{sec:num1}

\begin{figure}
\begin{center}
        \includegraphics[width=\columnwidth]{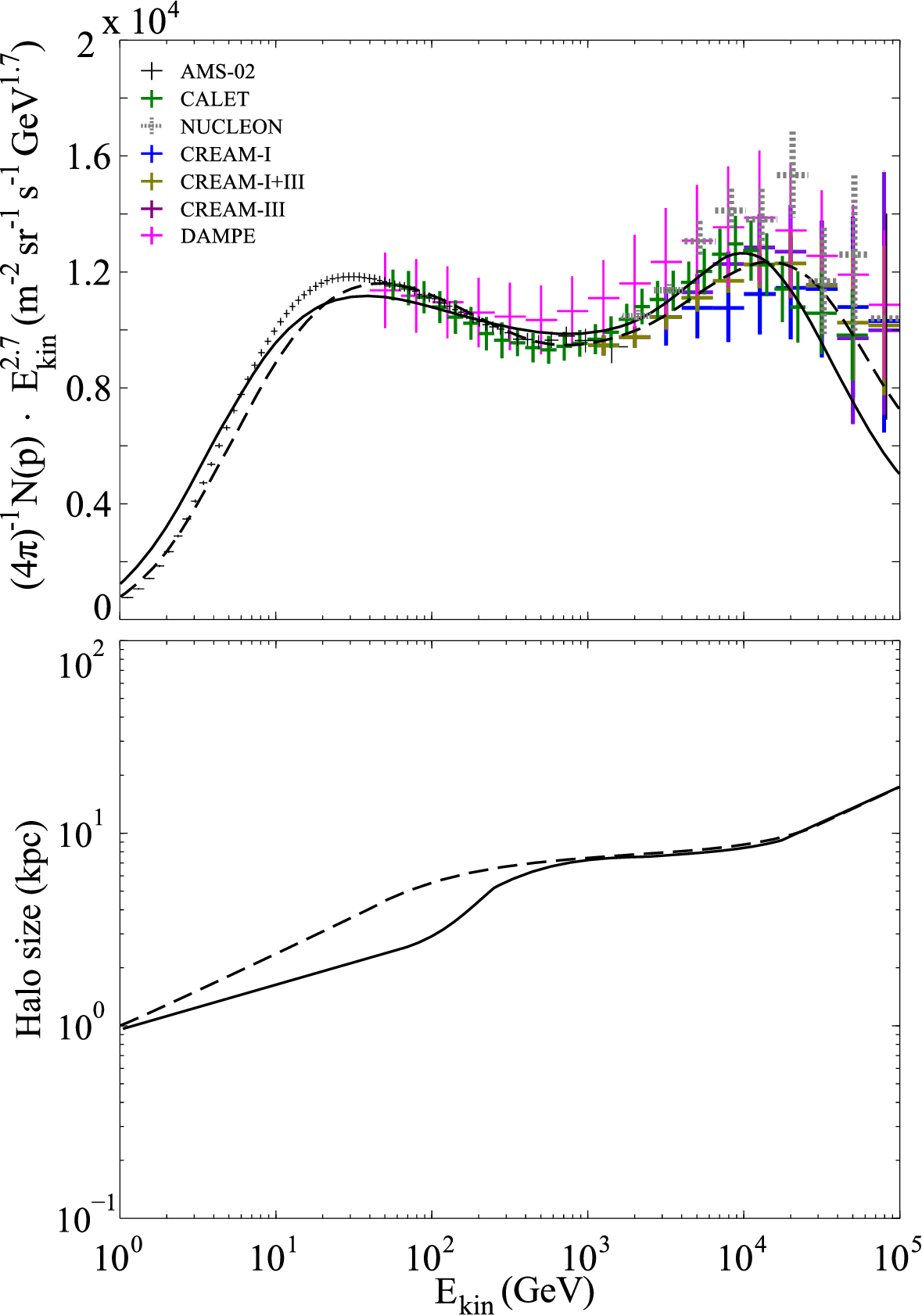}
    \caption{Energy spectrum of CR protons (upper panel) and energy-dependent halo size (lower panel), derived from the self-consistent model by \citet{chernyshov22} without (dashed line) and with (solid line) taking into account the ion-neutral damping. The symbols in the upper panel depict the observational data.}
    \label{fig:crp}
\end{center}
\end{figure}

\begin{figure*}[t]
\includegraphics[width=\linewidth]{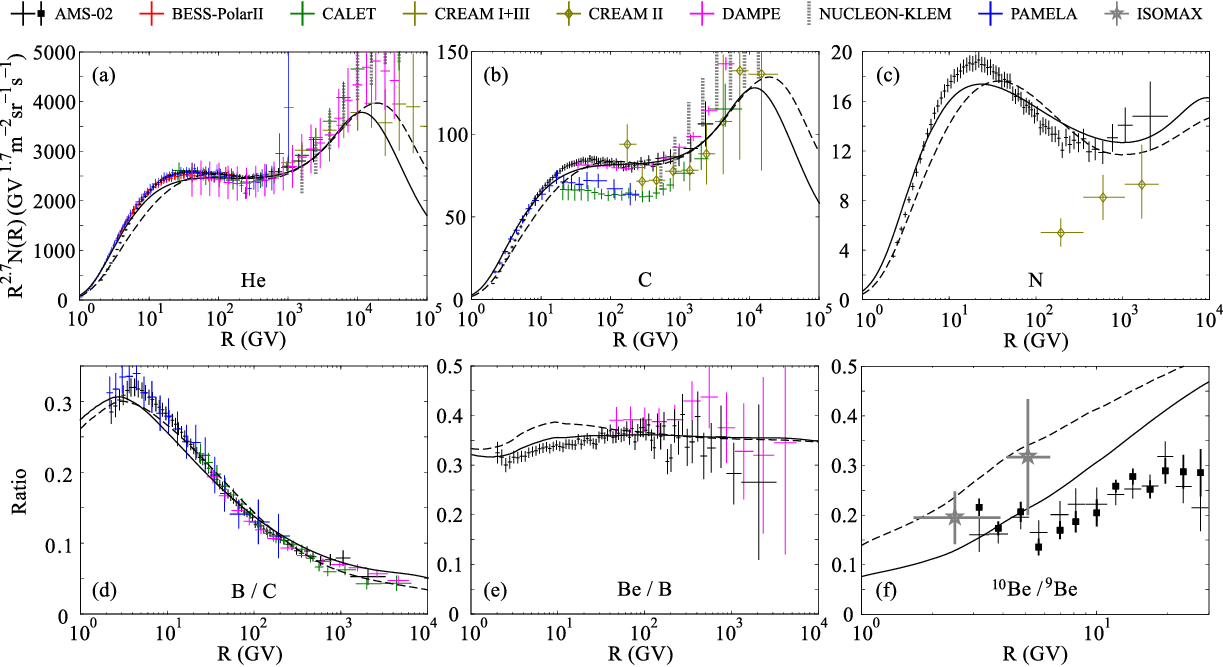}
\caption{Spectra of nuclei and their ratios versus rigidity. Top panels: Spectra of primary CR nuclei of helium (a), carbon (b), and nitrogen (c). Bottom panels: Abundance ratios of stable secondary to primary nuclei, represented by the boron-to-carbon ratio (d), as well as of unstable to stable secondary nuclei, represented by the beryllium-to-boron (e) and the beryllium-10-to-beryllium-9 (f) ratios. Our numerical results without and with taking into account the ion-neutral damping are plotted, respectively, by the dashed and solid lines. The symbols depict the observational data: preliminary AMS-02 data for the $^{10}$Be/$^9$Be ratio \citep{derome21} and their recent update \citep{Wei2023} are represented in the bottom right panel by the black crosses and squares, respectively.}
\label{fig:nuc_spectra}
\end{figure*}

We employ the self-consistent halo model by \citet{chernyshov22} with a two-component vertical distribution of ionized gas, which provides the best agreement with observations. The two components include a warm-ionized-medium (WIM) phase $n_{\rm WIM}(z) = n_{\rm WIM}^{(0)} \exp( - z / z_{\rm WIM})$ with constant temperature $T_{\rm WIM}$ and a hot-gas phase $n_{\rm hot}(z) = n_{\rm hot}^{(0)} \exp( - z / z_{\rm hot})$ with constant temperature $T_{\rm hot}$.

The vertical distribution of neutral gas in the disk is taken from GALRPOP at the galactocentric radius of 8~kpc, and we multiplied it by a numerical factor $\chi_n$ of about unity, which is needed to fit the B/C ratio. This factor accounts for the effect of diffusion along the Galactic disk. The vertical magnetic field is considered to be constant.In the injection term of primary nuclei, Equation~(\ref{eq:prim_CR_source}), each normalization constant $C_i$ is adjusted to fit observational data. We assume $\gamma = 2.26$ for the spectral index of all heavier nuclei, which is slightly different from that of protons, $\gamma_p = 2.32$. For a given set of model parameters, possible variations in $\gamma$ and $\gamma_p$ are limited by error bars for the observational data: according to \citet{Aguilar21}, $\Delta \gamma$ assessed for helium is $\ll 0.05$ while $\Delta \gamma_p\sim0.01$. 

We note that the density factor $\chi_n$ is a global parameter of the model. It affects all nuclei (including low-energy protons via continuous losses), and therefore iterative runs are needed to compute the spectra. The first run is made with $\chi_n = 1$. This allows us to assess the required correction by comparing the measured B/C ratio with the computed ratio, and use the updated value of $\chi_n$ for the next run. Since the resulting value of $\chi_n$ is close to unity, we do not perform further iterations.

To provide the best fit to the new data on proton spectrum reported by CALET \citep{adriani22}, we slightly adjusted parameters of our original paper \citep{chernyshov22}. For the vertical magnetic field in the halo we keep using the value of $B = 1~\mu$G, according to the order-of-magnitude estimates by \citet{Jansson2012}. The updated parameters of two-component ionized gas, based on observations reported in \citet{Ferriere1998} and \citet{Gaensler2008}, are $n_{\rm WIM}^{(0)} = 0.1$ cm$^{-3}$, $z_{\rm WIM}=0.4$ kpc, $T_{\rm WIM} = 0.7$ eV, $n_{\rm hot}^{(0)} = 10^{-3}$ cm$^{-3}$, $z_{\rm hot} = 2$ kpc, and $T_{\rm hot} = 200$ eV.  The resulting proton spectrum is plotted in the upper panel of Figure~\ref{fig:crp} by the dashed line together with the observational data, taken from AMS-02 \citep{agu15}, CALET \citep{calet19}, NUCLEON-KLEM \citep{nucleon19}, CREAM-I \citep{cream11}, CREAM-I+III \citep{cream17}, and DAMPE \citep{an2019}. The data are collected using Cosmic-Ray DataBase (CRDB v4.0) by \citet{Maurin2020}. The lower panel shows the computed halo size, set at the height where $\eta(R,z)=1$ \citep{chernyshov22}.

The results of the numerical solution of Equations (\ref{eq:main_iso}) for the above set of parameters are represented in Figure~\ref{fig:nuc_spectra} by the dashed lines. The observational data are taken from AMS-02 \citep{Aguilar16, Aguilar21}, BESS-PolarII \citep{Abe2015}, CALET \citep{Adriani23, Adriani23a}, CREAM-II \citep{ahn2009}, CREAM-I+III \citep{cream17}, DAMPE \citep{Alemanno21, dampe22, Parenti23}, NUCLEON-KLEM \citep{nucleon19}, and PAMELA \citep{Adriani11, Adriani14} using Cosmic-Ray DataBase (CRDB v4.1) by \citet{Maurin2020} and \citet{Maurin2023}. For $^{10}$Be/$^9$Be, we use preliminary AMS-02 data reported by \citet{derome21} and their recent update by \citet{Wei2023}, as well as ISOMAX data \citep{Hams2004}, all available only for relatively low rigidities. To account for the solar modulation, we use the force-field approximation with potential $\phi = 0.5$ GV \citep[][]{gleeson68}. Our self-consistent model shows reasonable agreement with the observational data for $R\gtrsim 30$~GV. To fit the B/C ratio, the density distribution from GALPROP is multiplied by the factor of $\chi_n = 0.85$.

We see that the agreement is also good for abundance ratios including unstable secondary nuclei, such as Be/B. However, for a pure unstable-to-stable ratio of $^{10}$Be/$^9$Be our results substantially overestimate the values deduced from preliminary AMS-02 data by \citet{derome21} and \citet{Wei2023}. Given that the B/C ratio (solely determined by $u_0(R)$ in our model) is well reproduced, the discrepancy in the $^{10}$Be/$^9$Be ratio may indicate that our model significantly underestimates $D_0(R)$ in the measured range of rigidities. 

\subsection{Effect of ion-neutral damping}

In Section~\ref{sec:num1}, we assumed that turbulence is solely regulated by nonlinear Landau damping. There are, of course, other damping mechanisms that may contribute, such as viscous and turbulent damping as well as ion-neutral damping \citep[see, e.g.,][]{kulsrud1969, xu2022}. However, in \citet{dogiel2020} we showed that viscous and turbulent damping can be safely neglected in the CR halo.

On the other hand, the ion-neutral damping may substantially affect the excitation-damping balance for MHD waves near the Galactic disk, suppressing the turbulence and, thus, increasing the value of diffusion coefficient $D_0(R)$. This effect can be included in the excitation-damping balance equation~(21) of \citet{chernyshov22} by adding the ion-neutral friction rate $\nu_{in}$ to the rhs of that equation. The energy density of MHD fluctuations is then described by their modified Equation~(22), indicating that turbulence at longer wavelengths can be completely damped. Since $\nu_{in}$ is proportional to the neutral gas density, the damping rapidly decreases with the height, and hence turbulence at larger $z$ remains practically unaffected.

To explore the effect of ion-neutral damping on the diffusion coefficient as well as on the outflow velocity and spectra of protons and nuclei, we first solve a set of self-consistent equations by \citet{chernyshov22} with the additional term $\nu_{in}$ in the excitation-damping balance. Then, we adjust the model parameters to be consistent with measurements. The best fit to the data in this case is achieved for a slightly lower temperature of the hot phase, $T_{\rm hot} = 170$~eV (instead of 200~eV) and a slightly larger scale height: $z_{\rm hot} = 2.3$~kpc (instead of 2~kpc). Also, the spectral index of proton sources is increased to $\gamma_p = 2.42$ (from 2.32) to compensate for a more efficient escape of low-energy particles. The resulting proton spectrum and the halo size are plotted in Figure~\ref{fig:crp} by the solid lines.

Next, we numerically solve Equations~(\ref{eq:main_iso}) using the computed diffusion coefficient and adjust the model parameters for nuclei sources. The spectral index of primary nuclei is now set to $\gamma = 2.37$ (instead of 2.26). The value of $\chi_n$ is adjusted as well, as the ion-neutral damping stimulates escape from the disk, which needs to be compensated by enhanced production of secondary nuclei. Therefore, the density factor to fit the B/C ratio is now $\chi_n = 1.15$ (instead of 0.85). The results are shown in Figure~\ref{fig:nuc_spectra} by the solid lines. One can see that the inclusion of ion-neutral damping makes the agreement with the observations noticeably better, in particular for the $^{10}$Be/$^9$Be ratio. 

It is worth noting that the difference between the solid and dashed lines at $R\gtrsim10^4$~GV, seen in Figures~\ref{fig:crp} and \ref{fig:nuc_spectra} for the primary CR spectra, can be eliminated if we also take into account a decrease of the magnetic field with the height (neglected above for simplicity). The lower panel of Figure~\ref{fig:crp} shows that the halo size exceeds $\sim10$~kpc for $R\gtrsim10^4$~GV, and  the expected field decrease at such heights should lead to the proportional reduction of $u_{\rm A}(z)$ (relative to the dependence assumed above); according to Equation~(\ref{eq:N_simple_stable}), this should proportionally increase the CR spectrum at the upper halo boundary and, thus, also boost the spectrum in the disk.

\section{Conclusions}

We estimated spectra of primary and secondary CR nuclei derived from the self-consistent halo model with nonlinear Landau damping by \citet{chernyshov22}. It is demonstrated that the model is able to reproduce observational abundance ratios of secondary to primary nuclei as well as of unstable to stable secondary nuclei. 

If the effect of ion-neutral damping of MHD waves near the Galactic disk is neglected, the model shows slight discrepancy between observational data and theoretical predictions at low rigidities (below $\sim30$~GV). Since the discrepancy is seen both in spectra of different nuclei and in their ratios, it cannot be removed by tuning the source term. Furthermore, the unstable-to-stable abundance ratio predicted by our model shows a factor of two excess over the values deduced for $^{10}$Be/$^9$Be from preliminary AMS-02 data by \citet{derome21} and \citet{Wei2023}. 

We show that agreement with observations at low rigidities can be noticeably improved, and, in particular, the discrepancy seen for the $^{10}$Be/$^9$Be ratio can be largely mitigated by adding the effect of ion-neutral damping, which suppresses the self-excited turbulence near the disk and, hence, leads to larger values of $D_0$. On the other hand, the discrepancy may also be attributed to other factors ignored in the model by \citet{chernyshov22}, such as CR-driven outflows from the disk, re-acceleration of CRs in the disk, influence of the disk on the turbulence in the halo, and the magnetic field structure in the halo (we assume the field lines to be vertical). A separate careful analysis is required in order to evaluate the actual importance of these factors, which is beyond the scope of the present paper.

Finally, we point out that the one-dimensional model considered cannot self-consistently take into account changes in the CR spectra across the Galactic disk, such as, for example, the spectral hardening toward the Galactic center \citep[see, e.g.,][]{yang2016, cerri2017}, as CRs are assumed to diffuse only in the vertical direction (both in the disk and in the halo). At the same time, since turbulence in our model is concentrated well above the disk (particularly when ion-neutral damping is taken into account), we expect the radial variations in the halo to be much smoother than those in the disk.

\begin{acknowledgements}
We are grateful to Laurent~Derome and Jiahui~Wei for allowing us to use preliminary AMS-02 data from \citet{derome21} and \citet{Wei2023}. We also would like to thank Lv~Xingjian for help with DAMPE data on helium nuclei, and the anonymous referee for the constructive suggestions.
\end{acknowledgements}

\bibliographystyle{aa}
\bibliography{refs2}

\end{document}